\documentclass[twocolumn,preprintnumbers,amsmath,amssymb]{revtex4-1}

\usepackage{graphicx}% Include figure files
\usepackage{dcolumn}% Align table columns on decimal point
\usepackage{bm}% bold math

\begin{document}

\title{Solitons in cavity-QED arrays containing interacting qubits}

\author{I-H.~Chen,$^{1}$  Y.~Y.~Lin,$^{1}$ Y.-C.~Lai,$^{1}$ E.~S.~Sedov,$^{2}$ A.~P.~Alodjants,$^{2}$ S.~M.~Arakelian,$^{2}$ and R.-K.~Lee$^{1}$}
\affiliation{
$^{1}$ Institute of Photonics Technologies, National Tsing-Hua University, Hsinchu, 300, Taiwan\\
$^{2}$ Department of Physics and Applied Mathematics, Vladimir State University named after A.G. and N.G. Stoletovs, Vladimir, 600000, Russia}

\date{\today}

\begin{abstract}
We reveal the existence of polariton soliton solutions in the array of weakly coupled optical cavities, each containing an ensemble of interacting qubits.
An effective complex Ginzburg-Landau equation is derived in the continuum limit  taking into account  the effects of cavity field dissipation and qubit dephasing.
We have shown that an enhancement of the induced nonlinearity can be achieved by two order of the magnitude with a negative interaction strength which implies a large negative  qubit-field detuning as well.
Bright solitons are found to be supported under perturbations only in the upper (optical) branch of polaritons, for which the corresponding group velocity is controlled by tuning the interacting strength. 
With the help of perturbation theory for solitons, we also demonstrate that the group velocity of these polariton solitons is suppressed by the diffusion process.
\end{abstract}
\maketitle

\section{Introduction}
To create new devices for quantum information storage and processing, macroscopically coherent coupled matter-light states are at the heart of researches, see e.g. \cite{Leuchs}.  Cavity quantum electrodynamics (cavity-QED), the study of coherent interactions between matters and quantized cavity electromagnetic fields, have provided a versatile and controllable platform to describe many interesting phenomena in this field; they are micro-masers, Purcell effect, squeezed state generation, atom trapping, and quantum state transfer, see e.g.~\cite{cavity-book, cavity-review}.
The simplest light-matter system, one cavity mode interacting with a single two-level atom, is described by the Jaynes-Cummings model \cite{JC}.
As the number of two-level atoms increases, collective effects, due to the field mediated interactions of atoms among themselves, can be described by the Dicke Hamiltonian and give rise to intriguing many-body phenomena \cite{Dicke}.
Apart from a single cavity configuration, cavity-QED arrays composed by engineered  optical cavities modes, few-level atoms and laser light modes, may serve as a many-body system for light \cite{JOSAB-review}.
Novel quantum phase transitions of light, from Mott-insulator, glassy, to super-solid states have been demonstrated to analyze critical quantum phenomena in conventional condensed matter systems by manipulating the interaction between photons and atoms \cite{Greentree06, Hartmann06, Angelakis06, Lei08}.

From practical point of view, two important prerequisites seem to be useful  to provide   quantum information processing and transmission. In particular, interactions are essential to generate necessary  quantum correlations and quantum entanglement \cite{Leuchs}. Second, weakly  coupled  cavity-QED arrays and lattices represent suitable candidates for quantum information transport by using photonic tunneling effect \cite{LongoSchmitteckert}. In particular, it is worth  noticing possible application of diamond NV centers for coupled cavity-QED arrays~\cite{ChunHsuSuGreentree}. The qubits based on NV centers  are potentially attractive because they are posing essentially longer dephasing time (up to few milliseconds) even at high enough temperatures \cite{LaddJelezko}.

Another comprehensive example on this way is the resonant excitation of a single quantum dot (QD)  strongly coupled with a photonic crystal nanocavity experimentally demonstrated in~\cite{EnglundMajumdar2010, HennessyBadolato2007, FaraonFushman2008}. Moreover, measurement of time-dependent second order autocorrelation function in such experiments  enables to distinguish photon blockade regime that is connected with anharmonic energy-level spacing and is, obviously, relevant to nonlinear effects in QD-light interaction in small volume cavities. Noticing that the detunings between frequency of coherent probe beam (or frequency of resonant exciton state), and  the bare-cavity frequency represent vital parameters in such experiments, cf.~\cite{EnglundMajumdar2010, FaraonFushman2008}.  
  
In the future such experiments permit to create on-chip coupled cavity-QED arrays with various architecture for few photon light field processing and transport integrated with telecom lines~\cite{EnglundFaraon2007}. It is known that quantum optical solitons are a natural candidate for quantum optical information prosessing.
Studies of  their transmission, formation, and transformation in coupled cavity-QED arrays represent  important steps for achieving this aim.
Polariton solitons have some advantages in this case. In particular, as demonstrated in \cite{SichKrizhanovskii2012}, polaritonic nonlinearity can be high enough in comparison with pure optical nonlinearities achieved, e.g. with pure optical cavity solitons in VCSELs~\cite{BarlandTredicce2002}, and can permit to achieve a soliton regime for essentially smaller particle (polariton) number.

More precisely, here we deal with the problem of  polariton soliton formation occurring at quantum matter-filed interface  in semiconductor quantum wells (QWs) embedded in Fabry-Perot microcavities. Nowadays polariton solitons and relevant superfluid behavior of non-equilibrium exciton-polaritons with narrow-band (GaAs) semiconductor structures are observed in a few labs,~\cite{SichKrizhanovskii2012,AmoPigeon2011}. The main physical features of such solitons are connected with the balance between dissipation effects and external pumping that occurs for parametrical processes allowing to create non-equilibrium exciton-polariton Bose-Einstein condensate (BEC) at lower polariton branch under the temperatures of few kelvins. The  solitons in experiments are excited within picosecond time scale and localized in micrometer-scale sizes.

Recently, in \cite{Sedov} we proposed a 2D polaritonic crystal model that represents cluster material exhibiting high nonlinear properties due to the small but macroscopical number of two-level atoms strongly coupled with optical field in the cavity lattice. Basing on the Holstein-Primakoff approach, we have shown that such nonlinearity can basically be created by means of two physical processes: that is a saturation effects occurring due to localization of macroscopically small number of atoms at each cavity, and atom-atom interaction on its own. Experimentally such a system can be designed by using 2D photonic crystal host with defect cavities~\cite{LeePainterKitzke2000}. Noticing that exciton-photon saturation effects can also be strong enough and accompanied by Coulomb repulsion processes for coupled cavity QED arrays based on the appropriate number of  multiple QWs structures,~cf.~\cite{EgorovSkryabinLederer2010}. Coherent matter-field interaction can be achieved in this case at room temperatures by using wide-band (ZnO) semiconductor microstructures,~cf.~\cite{TienChangLuYingYuLai}.

In the paper we consider the problem of soliton formation in 1D weakly coupled cavity arrays containing two-level systems (atoms, QDs, etc) that interact with single mode optical cavity fields. Instead of non-interacting (free) particles, several systems support interacting components which are referred to the qubit for simplicity.
For the ensemble of interacting qubits, the Lipkin-Meshkov-Glick (LMG) model is an example of the system producing maximal pairwise entanglement under the phase transition of its ground state \cite{Lipkin1965p188}. Recently, it has been proposed that the LMG model, originated in nuclear physics, may describe Josephson effect or two-mode Bose-Einstein condensate (BEC) \cite{Vidal2004p022107}.
In particular, relevant nonlinear model which is called in \cite{LeggettRMP2001} as "canonical Josephson Hamiltonian"  evokes great interest in the framework of efficient  spin squeezing \cite{KitagawaKitagawa1993} and resent experiments with  high precision measurements \cite{BodetEsteveOberthaler2010}.
Besides, the schemes for realizing a dissipative LMG model in optical cavity-QED \cite{Morrison2008p040403} and in circuit-QED \cite{Larson2010p54001} have been discussed. 
By introducing the interaction with a quantized optical cavity mode, the ground state phase transition \cite{Chen2007p40004}, maximal shared bipartite concurrence of the ensemble \cite{Gorlitz}, and dynamics from Rabi to Josephson oscillations \cite{Tsomokos2010, Blas} for the extended Dicke-LMG (DLMG) model have been studied. 

In this work, we extend DLMG model to the array of quantized field cavity modes and study nonlinear dynamics in such a qubit-cavity-QED array system.
In the continuum limit,  we use a \textit{multiple-scale envelope function} (MSEF) method~\cite{MSEF, cavity-array} to construct soliton solutions for considering nonlinear interactions among qubits as well as  dissipation effects for them. 
In particular, we derive and examine a complex Ginzburg-Landau equation (GLE) that supports "slow'' soliton propagations in the presence of polariton formation.
Being a step in the studies of collective properties of light with interacting media,
our results in the proposed qubit-cavity-QED system are shown to be connected with the formation of coupled matter-field  states, upper (UB) and lower (LB) branch polaritons, which would be natural carriers for quantum information processing at matter-field interface.  

\begin{figure}
\centering
\includegraphics[width=8.4cm]{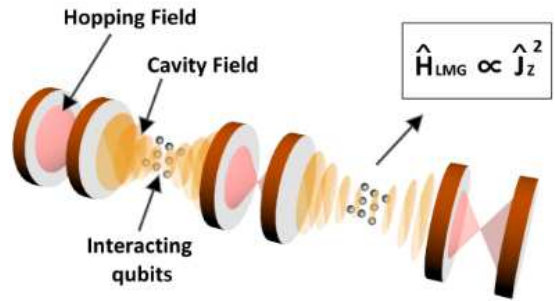}
\caption{\label{schematicdrawing} (Color online) Schematic for our proposed 1D qubit-cavity QED arrays, in which each cavity contains an ensemble of qubits with the interaction Hamiltonian among them  described by the LMG model, $\hat{H}_\text{LMG} \propto \hat{J}_z^2$. }
\end{figure}

\section{\label{mofelofinteracting} MODEL OF INTERACTING QUBIT-CAVITY-QED ARRAYS}

As illustrated in Fig.~\ref{schematicdrawing} we consider a one-dimensional (1D) array of optical cavities, each containing an ensemble of  $N_{i} $ of interacting qubits. The interaction inside each single cavity is described by  Dicke-LMG model Hamiltonian, $H_{i}^{{\rm DLMG}} $, and is represented as, cf.~\cite{LeggettRMP2001,KitagawaKitagawa1993,BodetEsteveOberthaler2010,Morrison2008p040403}
\begin{eqnarray} \label{eq:ExtDicke}
\hat{H}_i^{\text{DLMG}} = \hbar \Delta_i \hat{a}_i^\dagger\hat{a}_i + \frac{\hbar \eta_i}{N_i} \hat{J}_{z, i}^{2} + \frac{\hbar g_i}{\sqrt{N_i}}  \left(\hat{a}_i  \hat{J}_{+,i} + \hat{a}_i^{\dagger}\hat{J}_{-,i} \right).\nonumber\\
\end{eqnarray}

Subscript $i$ labels the $i$-th cavity to be used afterwards. 
Detuning frequency $\Delta_i$ has been defined as the difference between qubit transition frequency $\omega_i$ and quantized field frequency $\omega_{f}$, i.e., $\Delta_i = \omega_i - \omega_f$; orbital angular momentum representation, $\hat{J}_{j, i}$ with $j=x,y,z,\pm$, has been chosen to describe the qubit ensemble, while the field is described by its creation (annihilation) operators, $\hat{a}_i^{\dagger}$ ($\hat{a}_i$); the field-ensemble and the inter-qubit coupling are represented by the constants $g_i$ and $\eta_i$.

Hamiltonian in Eq.~(\ref{eq:ExtDicke}) is equivalent to the LMG model in the limit $g_i = 0$, and hereby is called Dicke-LMG Hamiltonian, which can be obtained from the Gross-Pitaevskii equation describing a two-species BEC (see~e.~g.~\cite{LeggettRMP2001}) interacting with a quantum field~\cite{Chen2007p40004}. 
Experimental realizations providing an assorted range of tunable parameters for the DLMG model may include a two-hyperfine-structure-defined-modes BEC coupled with a quantum cavity field mode through one microwave photon process; e.g. trapped hyperfine ground states of a Sodium BEC inside a microwave cavity~\cite{Gorlitz}. 
In the area of solid state physics, superconducting qubit devices seem to be promissing candidates to realize qubit-cavity-QED being under discussion, see~e.~g.~\cite{DevoretWallraffMartinisARX}. In particular, arrays interacting superconducting qubits coupled with the quantum field mode of a coplanar waveguide resonator may be provided the ensemble sizes are small~\cite{Tsomokos2010}. 

For the configuration of an array of cavities, the qubits are assumed to be confined inside the cavities, while photons can travel from one cavity to another. We also assume that the traveling process for photons is dominated by the hopping between adjacent sites. The value of hopping matrix elements is a function of the distance between cavities. If the distance between each cavity is sufficiently small, a cavity field will only hop to the nearest neighbors. Then the total Hamiltonian of interaction for this qubit-cavity-QED array is given as~(cf.~\cite{Greentree06}), 
\begin{eqnarray} \label{eq:TotalHam}
\hat{H}_I&=& \sum_{i=1}^{M} [\hat{H}_i^{\text{DLMG}} -\hbar \alpha \left( \hat{a}_i^\dagger \hat{a}_{i+1} + \hat{a}_{i+1}^\dagger \hat{a}_i \right)], 
\end{eqnarray}
where $\alpha$ is a photon-hopping strength from cavity site $i$ to $i+1$, and $M$ counts the number of cavities. In the present work, we assume that only one species of qubits is considered and that the number of ensemble qubits is the same in each cavity, i.e., $\eta_{i} \equiv \eta $, $\omega _{i} \equiv \omega $, and $N_{i} \equiv N$. For simplicity, the parameters are taken as real and homogeneous so that the detuning frequency, hopping and coupling strengths be all identical in each cavity site, i.e., $\Delta _{i} \equiv \Delta $, $g_{i} \equiv g$, and $\alpha _{i} =\alpha $, respectively.
Physically, this approach is valid if the separation of two adjoint cavities is macroscopically large in comparison with the atomic de Broglie wavelength and overlapping of  atomic wave functions from neighbor cavities is negligibly small. 

If most of qubits stay in the ground state, we can apply Holstein-Primakoff transformation~\cite{LaurentPatrickLevy} by mapping qubit excitation operator $\hat{J}_{j,i}$ on the Schwinger representation~\cite{Schwinger} for a two-level system, i.e., $\hat{J}_{+, i} = \sqrt{N} \hat{b}_i^\dagger \left(1 - \hat{b}_i^\dagger \hat{b}_i/N\right)^{1/2}$ and $\hat{J}_{z, i} = \hat{b}_i^\dagger \hat{b}_i -N/2$, along with the commutation relation $[\hat{b}_i, \hat{b}_j]=\delta_{ij}$.  Practically, the number of qubits $N$ is large enough and one can expand the orbital angular momentum representation for the qubit ensemble up to the order of $1/N$, i.e., $[1 - (\hat{b}_i^\dagger \hat{b}_i / N)]^{1/2} \approx 1- \hat{b}_i^\dagger \hat{b}_i/2N$. We are working within the so-called low-excitation-density limit when the qubits mostly populate their ground levels.

Since a photonic field is  an order parameter for the system described by Hamiltonians (\ref{eq:ExtDicke}), (\ref{eq:TotalHam}) it is non-zero for superfluid phase of coupled matter-field state in the presence of photon tunneling between neighbor cavities, see e.g.~\cite{Greentree06, Barinov}. At the same time  low-excitation-density limit implies that the average photon number at each cavity should be essentially  smaller than the average number of qubits at the excited state, that is $\left\langle \hat{a}_{i}^{\dag} \hat{a}_{i} \right\rangle \ll \left\langle \hat{b}_{i}^{\dag } \hat{b}_{i} \right\rangle$ -- cf.~\cite{Barinov}.

Now, since number $N$ is assumed to be large enough,  we are able to neglect higher-order terms in the expansion of $\hat{J}_{+,i} $ operator. Then, we can transform the Hamiltonian in Eq.~(\ref{eq:TotalHam}) into 
\begin{eqnarray}\label{eq:TotalHam2}
\hat{H}_{\text{HP}} &=& \hbar\sum_{i=1}^M \{-  \alpha \left( \hat{a}_i^\dagger \hat{a}_{i+1} + \hat{a}_{i+1}^\dagger \hat{a}_i\right) + g \left(\hat{a}_i^\dagger \hat{b}_i + \hat{a}_i \hat{b}_i^\dagger \right) \nonumber\\
&-&\frac{g}{2N}\left[\hat{a}_i^\dagger \hat{b}_i^\dagger \hat{b}_i \hat{b}_i + \hat{a}_i \hat{b}_i^\dagger\hat{b}_i^\dagger\hat{b}_i\right]\nonumber\\
&+&\frac{\eta}{N}\left(\hat{b}_i^\dagger \hat{b}_i \hat{b}_i^\dagger \hat{b}_i -N \hat{b}_i^\dagger \hat{b}_i \right) + \Delta \hat{a}_i^{\dagger} \hat{a}_i \}.
\end{eqnarray}

Based on this Hamiltonian, Heisenberg's equations of motion for the involved two bosonic operators of cavity modes $\hat{a}_{i} $ and qubit excitation $\hat{b}_{i} $ are
\begin{eqnarray}
i \frac{\partial}{\partial t} \hat{a}_i 
 &=&  \Delta \hat{a}_i -\alpha \left(\hat{a}_{i+1}+ \hat{a}_{i-1}\right)+ g\left[ \hat{b}_i - \frac{1}{2 N}\left(\hat{b}_i^\dagger \hat{b}_i \hat{b}_i\right)\right] ,\nonumber \\ \label{eq:Heisenberg_1}\\
i \frac{\partial}{\partial t} \hat{b}_i &=& g\left[ \hat{a}_i - \frac{1}{2 N}\left( \hat{a}_i^\dagger \hat{b}_i \hat{b}_i + 2 \hat{a}_i \hat{b}_i^\dagger \hat{b}_i\right)\right] -\eta \hat{b}_i + \frac{2\eta}{N} \hat{b}_i^\dagger \hat{b}_i \hat{b}_i. \nonumber \label{eq:Heisenberg_2}\\  
\end{eqnarray}

Here, we have neglected quantum fluctuations due to a large number of qubits considered.

To have mean-field solutions,  we replace the pair of operators $( \hat{a}_{i}, \hat{b}_{i} )$ by the corresponding expectation values $(\psi _{i} ,\beta _{i} )$, and approximate this array configuration to a continuous model, i.e., $\psi _{i+1} +\psi _{i-1} \approx 2\psi (x,t)+d^{2} \frac{\partial ^{2} }{\partial x^{2} } \psi (x,t)$, with the distance between two adjoint optical cavities denoted by $d$.
By considering the conservation of the total photon number and qubit excitations, we can re-normalize the variables with respect to the number, $N_{{\rm pol}} $, defined as $N_{{\rm pol}} =|\psi |^{2} +|\beta |^{2}$. The conserved quantity $N_{\rm pol}$ is also known as the number of polaritons at each cavity~\cite{Chestnov}.
By replacing the cavity field and qubit excitation by $\psi \to \psi /\sqrt{N_{{\rm pol}} } $ and $\beta \to \beta /\sqrt{N_{{\rm pol}} } $, respectively, now we have a normalization condition for the cavity field and qubit excitation, $|\psi|^2+|\beta|^2= 1. $

Thus, in this continuum limit, the equations of motion in Eqs.~(\ref{eq:Heisenberg_1},\ref{eq:Heisenberg_2}) become, 
\begin{eqnarray}
i \partial_t \psi &=& (\Delta -2\alpha -i\gamma _{c} )\psi -\alpha d^{2} \partial _{xx} \psi +g\beta - U_{\rm sat}|\beta |^{2} \beta, \nonumber \\ \label{eq:HeisenbergValue_1}\\
i\partial _{t} \beta &=& -(i\Gamma _{d} +\eta )\beta +g\psi \nonumber \\
&-& U_{\rm sat} \left[\beta ^{2} \psi ^{*} +2\psi |\beta |^{2} \right]+ U_{\rm int} |\beta |^{2} \beta, \label{eq:HeisenbergValue_2}
\end{eqnarray}
where $U_{\rm int} \equiv 2\eta  n_{\rm pol}$, $U_{\rm sat} \equiv \frac{g n_{\rm pol}}{2}$; $n_{\rm pol} = N_{\rm pol}/N $ is a polariton number density.
In Eqs.~(\ref{eq:HeisenbergValue_1},\ref{eq:HeisenbergValue_2}) we have a phenomenologically introduced decay rate for photonic field $\gamma _{c} $ that characterizes the leakage of photons in the cavity and dephasing rate $\Gamma _{d}$ for the qubit system.
Physically, parameter $U_{\rm sat}$ characterizes a nonlinear saturation effect under the qubit-light interaction. Parameter $U_{\rm int}$ is responsible for qubit-qubit interactions within the LMG model.
Let us briefly discuss the applicability of our model (Eqs.~(\ref{eq:HeisenbergValue_1},\ref{eq:HeisenbergValue_2})) for different physical systems representing qubits.

\textit{Example~1: cavity QED with atomic qubits.} Here we discuss cavity QED array with two-level atoms as a qubit system.  To be more specific, we consider ultracold two-level rubidium atoms with resonance frequency $\omega _{ab} /2\pi =382{\rm THz}$ that corresponds to mean weighted rubidium  D-lines \cite{Chestnov}. The atomic polarization dephasing rate and the minimal value of each cavity field decay rate can be taken as several tens of megahertz's that corresponds to cavity quality factor  $Q=\omega _{ab} /2\gamma _{c} \sim 10^{6} $. The strength of interaction of a single atom   with a quantum optical field is taken as $g_{0} /2\pi =89.5{\rm MHz}$ at each cavity with the effective  volume of atom-field interaction $V=5000{\rm \mu m^{3} }$. To achieve a strong atom-field coupling regime, see Eq. (\ref{couplingcondition}) below, one can propose a macroscopically large number of atoms at each cavity, say $N=5\times 10^{5} $. This number is relevant to the density $\rho =10^{14} {\rm cm^{-3} }$ of ultracold atoms that implies a collective atom-filed coupling parameter as $g\equiv g_{0} \sqrt{N} =2\pi \times 63.2{\rm GHz}$ at each cavity.  At the same time the parameter that describes atom-atom interaction $\eta ={4\pi \hbar a_{sc} \rho \mathord{\left/ {\vphantom {4\pi \hbar a_{sc} \rho  m}} \right. \kern-\nulldelimiterspace} m} $ in the Born approximation  can be estimated as  $\eta =2\pi \times 0.73{\rm kHz}$, where $a_{sc} $ is atomic scattering length that we choose as $a_{sc} =5{\rm nm}$, cf.~\cite{ChinGrimm2010}. Thus, for atomic QED array ${U_{\rm int} \mathord{\left/ {\vphantom {U_{int}  U_{sat} }} \right. \kern-\nulldelimiterspace} U_{sat} } \simeq 10^{-8} $ and  we can neglect atom-atom interaction processes in Eqs.~(\ref{eq:HeisenbergValue_1},\ref{eq:HeisenbergValue_2}). 

\textit{Example 2: cavity QED with excitonic qubits in QWs.} A 2D version of Eqs.~(\ref{eq:HeisenbergValue_1},\ref{eq:HeisenbergValue_2}) describes properties of exciton-polaritons that can be created in a microcavity with Bragg mirrors under strong coupling between excitons  confined in QWs and a cavity photonic mode, see e.g.~\cite{EgorovSkryabinLederer2010,KarrBaas2004}.
It has been shown recently that the strong coupling condition can be  achieved in the system of   photonic crystal polaritons ~\cite{Baioni2009}, which can be a possible platform to implement qubits being proposed in a cavity-QED-array.
The typical values of Rabi splitting frequency for QWs based on GaAs or CdTe is about few of meVs. The saturation term  $U_{sat} $ in Eqs.~(\ref{eq:HeisenbergValue_1},\ref{eq:HeisenbergValue_2}) for QWs structure can be represented as $U_{sat} \simeq {g\mathord{\left/ {\vphantom {g n_{sat} S}} \right. \kern-\nulldelimiterspace} n_{sat} S} $, where  $n_{sat} $ is exciton saturation density, $S$ is a quantization area.  The typical values for  $n_{sat} $ and   $U_{sat} $ are  $1.4\times 10^{11} {\rm cm}^{{\rm -2}} $ and $2\pi \times 0.65{\rm MHz}$, respectively, taken for exciton Bohr radius $10{\rm nm}$ and excitation laser spot $S\simeq 400{\rm \mu m}^{{\rm 2}} $.  The exciton-exciton interaction term in ~(\ref{eq:HeisenbergValue_2}) is evoked by Coulomb interaction and about two order times larger than $U_{sat} $, cf.~\cite{KarrBaas2004}. Thus, we can neglect exciton-photon saturation effects for excitonic qubits. Typical experimentally accessible  values  for relaxation processes in semiconductor structures can be approached as $\Gamma _{d} \simeq 2\pi \times 12.1{\rm GHz}$,  $\gamma _{c} \simeq 2\pi \times 50{\rm GHz}$ that implies exciton-polariton lifetime within tens of picoseconds. 

The set of coupled nonlinear equations~(\ref{eq:HeisenbergValue_1},\ref{eq:HeisenbergValue_2}) is the starting point of our work, and soliton solutions both for the wave-packet envelope of cavity field $\psi $ (order parameter) and for qubit excitations $\beta $ are considered analytically below.

\section{\label{gleforpolaritonqubit} Dispersion relations and group velocities}
To construct solitons in this cavity-QED arrays with interacting qubits, we seek for wave-packet solutions by the \textit{multiple-scale envelope function} method ~\cite{MSEF, cavity-array}. With the introduction of different length and time scales, i.e., $x_{m } =\lambda ^{m} x$ ($\lambda \ll 1,m=0,1,2\ldots $) and $t_{m} =\lambda ^{m} t$ ($\lambda \ll 1,m=0,1,2\ldots $), we can expand photon field, $\psi $, and qubit excitation, $\beta $, as 
\begin{eqnarray*}
\psi &=& \lambda \psi^{(1)}+\lambda^2 \psi^{(2)}+\lambda^3 \psi^{(3)}+\ldots \\
\beta &=& \lambda \beta^{(1)}+\lambda^2 \beta^{(2)}+\lambda^3\beta^{(3)}+\ldots
\end{eqnarray*}
By substituting these expansions in Eqs.~(\ref{eq:HeisenbergValue_1}-\ref{eq:HeisenbergValue_2}), we can gather all terms that are proportional to the first order of $\lambda $ and find, 

\begin{eqnarray}
i\partial_{t_0} \psi^{(1)} &=& (\Delta -2\alpha -i\gamma _{c} ) \psi^{(1)} - \alpha d^2 \partial_{x_0}^2\psi^{(1)} + g\beta^{(1)}, \nonumber \\ \label{eq:Heisenberg1stAppr_1}\\
i\partial_{t_0}\beta^{(1)}&=& -(i\Gamma_{d} +\eta )\beta ^{(1)} +g\psi ^{(1)}. \nonumber \\ \label{eq:Heisenberg1stAppr_2}
\end{eqnarray}

This first-order expansion for the qubit excitation $\beta^{(1)}$ in Eq.~(\ref{eq:Heisenberg1stAppr_2}) supports a plane wave solution, from which we can find a corresponding dispersion relation by using the solution in the form of  $\psi ^{\left(1\right)} =E^{(1)} e^{i\left(kx_{0} -\omega t_{0} \right)} $ and  ${\beta ^{\left(1\right)} =\frac{g}{\omega +\eta +i\Gamma_{d} } E^{(1)} e^{i\left(kx_{0} -\omega t_{0} \right)}} $. The corresponding carrier frequencies are
\begin{eqnarray}
&&\omega \equiv  \omega _{\pm } =\frac{1}{2} \left[\Delta -\eta -2\alpha \left(1-\frac{d^{2} k^{2} }{2} \right)-i\left(\gamma _{c} +\Gamma_{d} \right) \right.  \nonumber \\
&&\left. \pm \left(\left(\eta +\Delta -2\alpha \left(1-\frac{d^{2} k^{2} }{2} \right)-i\left(\gamma _{c} -\Gamma_{d} \right)\right)^{2} +4g^{2} \right)^{1/2} \right],\nonumber \\ \label{Eq:DispRel}
\end{eqnarray}
where $k$ is a wave number.  
Physically, Eq.~(\ref{Eq:DispRel}) reproduces a familiar result for frequencies of two branches of polaritons~\cite{PauBjork},  which are denoted as UB ($\omega_+$) and LB ($\omega_-$), respectively.  

It is important to note that field ($\psi ^{(1)}$) and qubit excitation ($\beta ^{(1)}$) variables become exact solutions of Eqs.~(\ref{eq:HeisenbergValue_1},\ref{eq:HeisenbergValue_2}) neglecting nonlinear effects, i.e., setting $U_{\rm sat}=U_{\rm int}=0$. Hence, it is easy to understand that MSEF method developed here is valid if dispersion characteristics of polaritons, see Eq. (\ref{Eq:DispRel}), cannot be essentially modified taking into account nonlinear effects. Strictly speaking, the condition
\begin{equation}
\label{UsatUint}
U_{\rm sat},U_{\rm int} \ll 2g
\end{equation}
should be fulfilled for coupled qubit-light systems described by Eqs.~(\ref{eq:HeisenbergValue_1},\ref{eq:HeisenbergValue_2}). In practice this condition implies achieving low-excitation density limit ${n_{\rm pol} \ll 1}$ discussed above.

Next, for the second order of multiple-scales, i.e., $\lambda ^{2} $, we can have a linear wave equation for a wave-packet  envelope 
\begin{eqnarray}
\label{WPEnvelope}
\left(\partial _{t_{1} } +v_{\pm } \partial _{x_{1} } \right)E_{}^{(1)} =0,
\end{eqnarray} 
from which one can find the corresponding group velocities, 
\begin{eqnarray}
\label{velocity}
v_{\pm} =\partial _{k} \omega _{\pm } =\frac{2\alpha kd^{2} \Omega _{\pm }^{2} }{\Omega _{\pm }^{2} +g^{2}} \end{eqnarray} 
that are defined for UB ($v_{+} $) and LB ($v_{-} $) polariton wave-packets. Below we consider variable $E^{(1)} $ in connection with Eq.~(\ref{WPEnvelope}) in the moving frames $\xi _{\pm } =x_{1} -v_{\pm } t_{1} $. In Eq. (\ref{velocity})  we have also introduced characteristic frequencies 
\begin{eqnarray}
\label{charfreq}
\Omega _{\pm } =\frac{1}{2} \left[\delta -i\gamma \pm \left(\left(\delta -i\gamma \right)^{2} +4g^{2} \right)^{1/2} \right],
\end{eqnarray}
where $\gamma =\gamma _{c} -\Gamma_{d} $ is an effective qubit-field decay rate for the system and 
$\delta =\Delta +\eta -2\alpha \left(1-{d^{2} k^{2} \mathord{\left/ {\vphantom {d^{2} k^{2}  2}} \right. \kern-\nulldelimiterspace} 2} \right)$ is a total \textit{momentum dependent} qubit-light detuning. 
Now, the qubit interaction parameter $\eta $ introduces an additional phase shift that can be used to tune the relative dispersion relations of polaritons.

It is noted that the group velocities $v_{\pm } $ defined in Eq. (\ref{velocity}) are complex numbers in general. It is known that this leads to the additional deformation of a pulse envelope propagated in the medium~\cite{Bukhman2001}.
The above MSEF method works only for small decay rates if a strong coupling qubit-light condition is valid, that is 
\begin{equation}
\label{couplingcondition}
\gamma _{c} ,{\rm \; }\Gamma_{d} \ll 2g.
\end{equation} 

In particular, in this limiting case the imaginary part of  frequencies  ${\rm Im}(\omega _{\pm } )$, which characterizes the decay of polaritonic systems,  is smaller than the real one ${\rm Re}(\omega _{\pm } )$, i.e., ${\rm Im}(\omega _{\pm } ) \ll {\rm Re}(\omega _{\pm } )$.
If the condition in Eq. (\ref{couplingcondition}) is not met, the group velocity can be correctly determined in terms of the energy flow in the medium only~\cite{Loudon1970}.  However, this is not the case of our present consideration. 

\begin{figure}
\includegraphics[width=8.4cm]{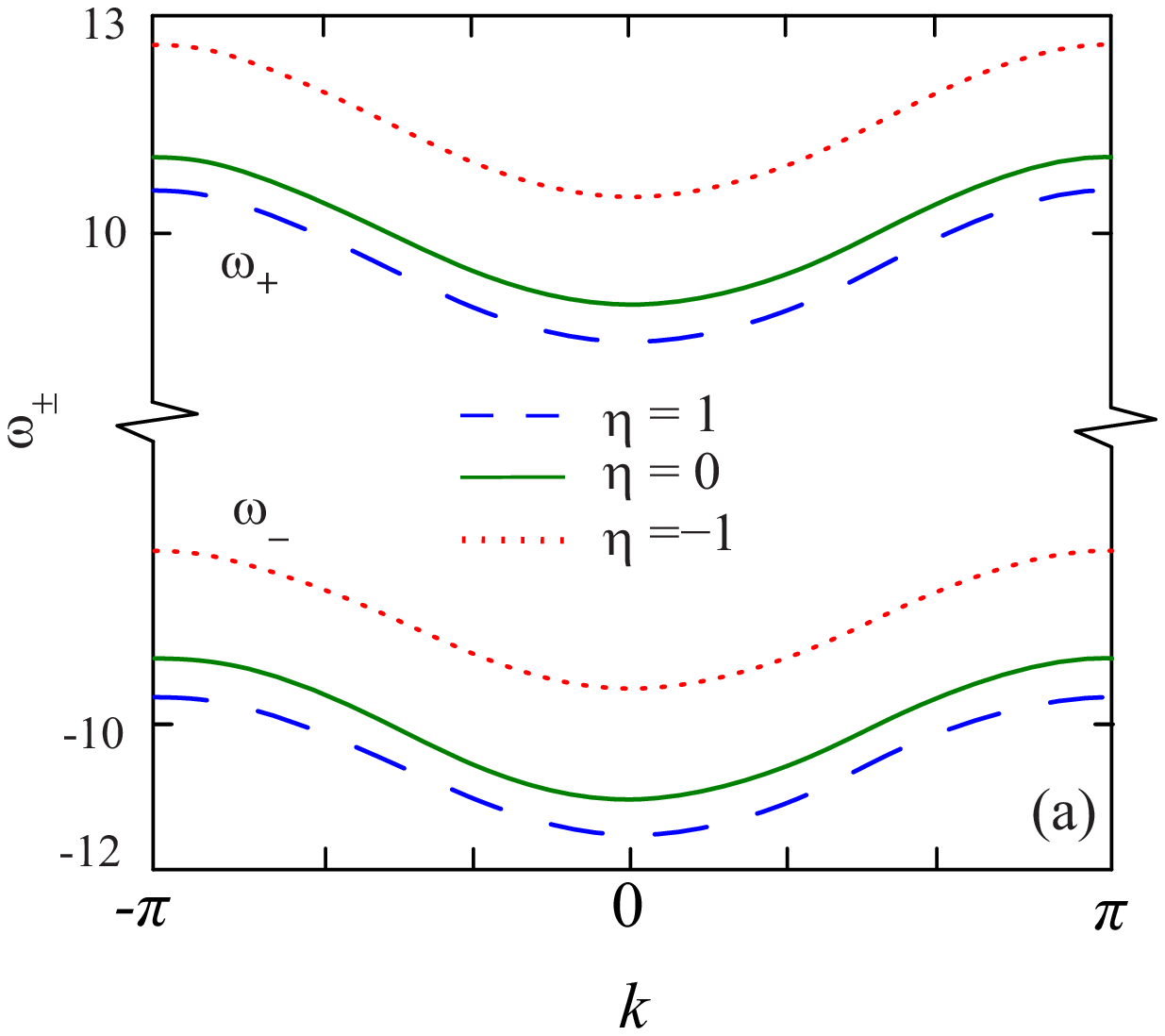}
\includegraphics[width=9cm]{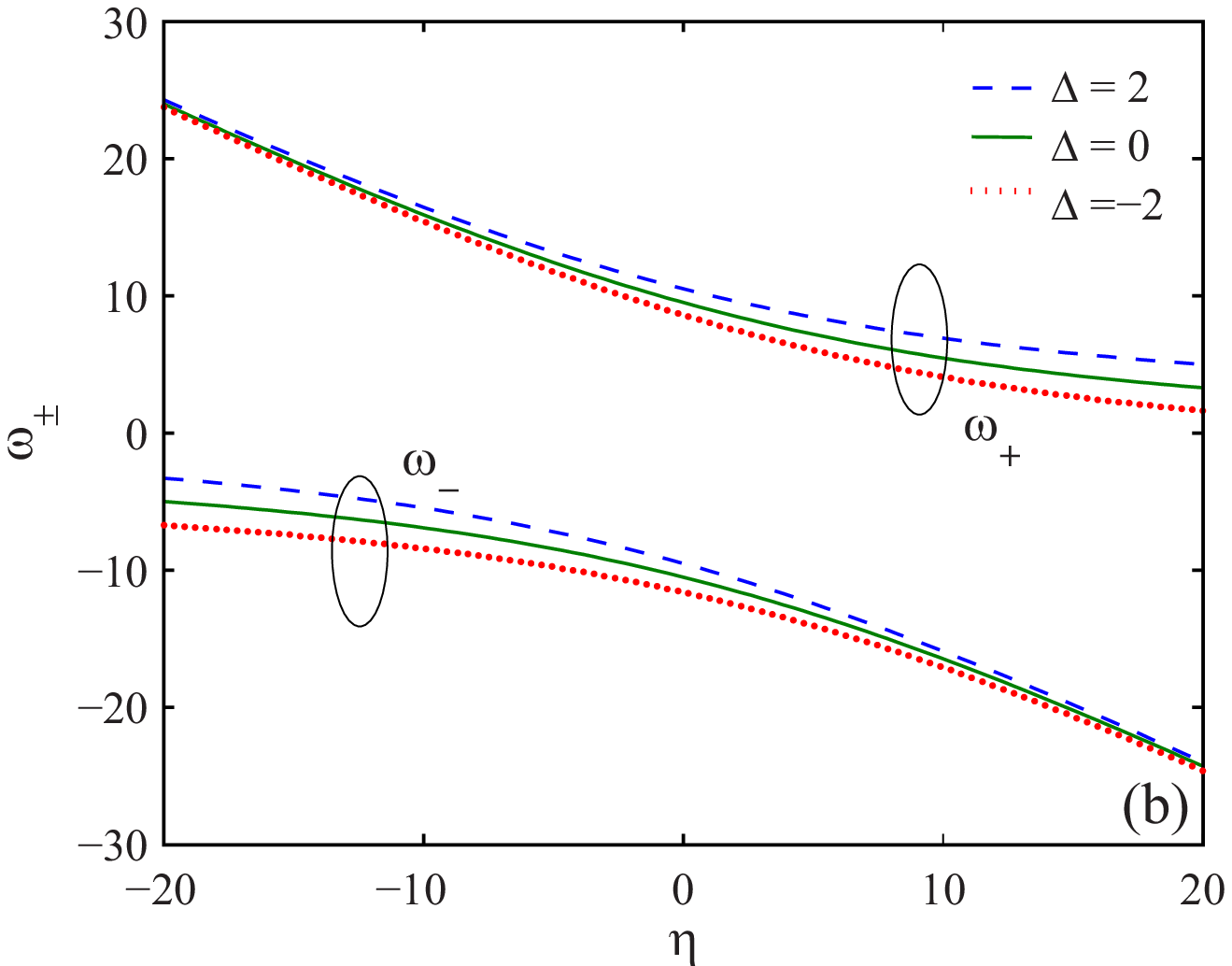}
\includegraphics[width=8cm]{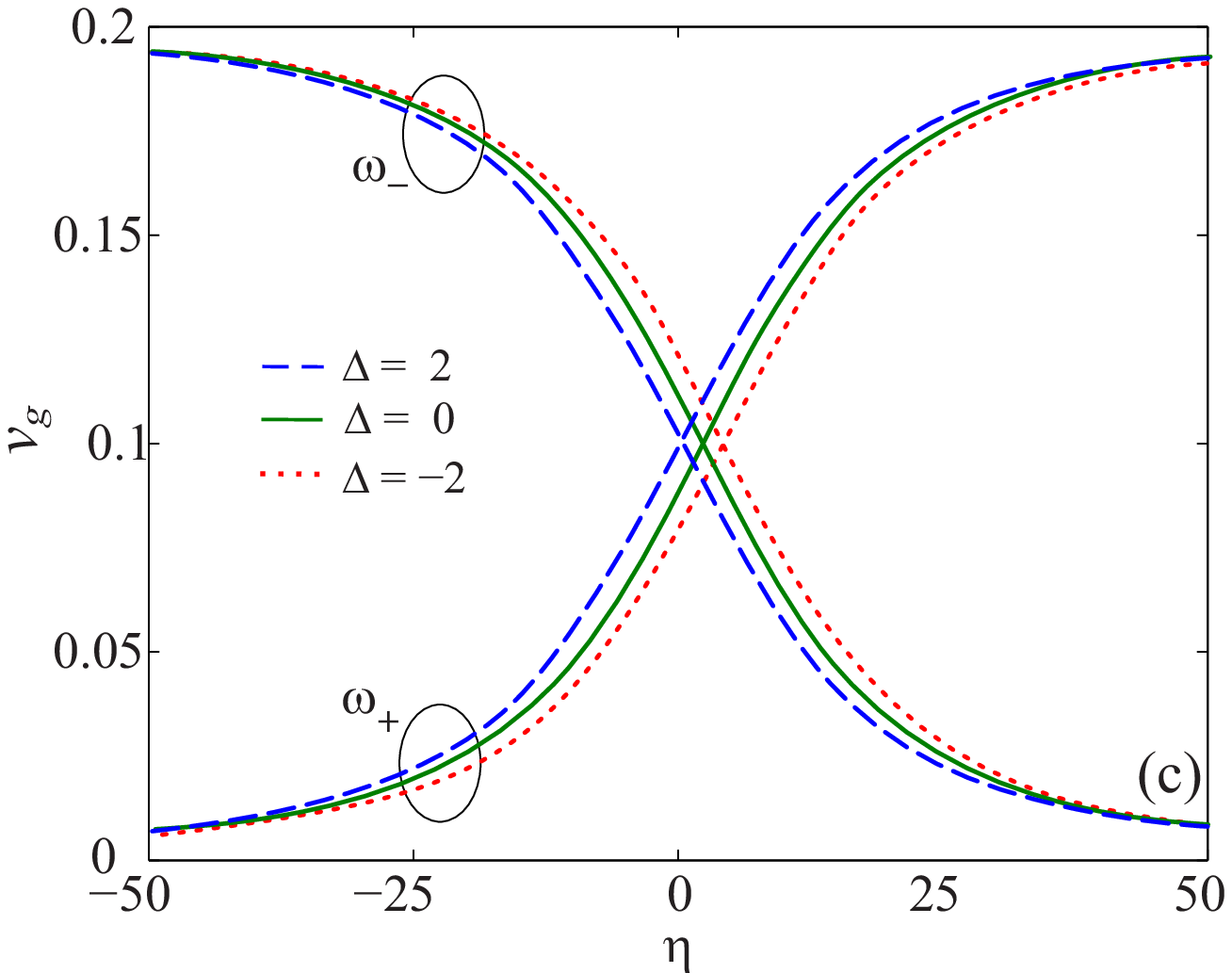}
\caption{(Color online) (a) Two branches in the dispersion relation, labeled as upper $\omega_+$ and lower $\omega_-$ branches, are shown as a function of  the wave number $k$ at the resonance condition $\Delta = 0$,  but for different qubit interaction strengths, $\eta = -1$, $0$, and $1$, respectively.
(b) The carrier frequencies $\omega_\pm$ and (c) the group velocity $v_{\pm}$ for the wave packet are  shown  as a function of interacting strength $\eta$ under the condition $k = 0.1$, $\gamma=0$, but with different frequency detunings, i.e., $\Delta = -2$, $0$, and $+2$, respectively.
Other parameters used are $\alpha=d=1$, $g=10$, and $n_{\rm pol}=0.01$.}
\end{figure}

With the introduction of the qubit interaction, in Fig. 2(a), we show the dependence of these two branches $\omega_\pm$ on the wave number $k$ under the resonance condition $\Delta = 0$, but for different qubit interaction strengths, $\eta = -1$, $0$, and $1$, respectively. Equation (\ref{Eq:DispRel}) is valid for a small quasi-momentum $k$, such as $kd \ll 1$. 
In particular, if we approximate $1-d^{2} k^{2} /2\approx \cos (kd)$ in Eq.~(\ref{Eq:DispRel}), then, by setting $\eta =0$ and $\gamma _{c} =\Gamma_{d} =0$, we can recognize true dispersion relation taken at first Brillouin zone for periodic structure, see Fig.~2 and cf.~\cite{Sedov}. 
For a small quasi-momentum  $k$ we can have parabolic dispersion relation that is inherent in LB polaritons occurring in semiconductor microcavities~\cite{SichKrizhanovskii2012, AmoPigeon2011}.
From Fig. 2(a), one can see that a negative value of $\eta$, accounting a repulsive interaction among the qubits, moves the two branches $\omega_\pm$ upward to a higher value; while a positive one moves the dispersion relation downward. 

By collecting terms of $\eta -\Delta$ in Eq.~(\ref{Eq:DispRel}) under the condition (\ref{couplingcondition}),  we can make the linearization approximation for the dispersion relation, i.e.,
\begin{eqnarray}
&& \omega_{\pm} \approx -\alpha \left( \cos (kd) + \frac{\eta - \Delta}{2 \alpha}\right) \pm \mid{}g{}\mid,
\end{eqnarray}
where one can view this qubit interaction as a linear frequency detunging under the conditions of ${2 \Delta \alpha \cos (kd) < \alpha ^{2} \left( \cos (kd) + \frac{\eta - \Delta}{2 \alpha}\right) ^{2} \ll \mid g \mid ^{2}}$.
In this case, we obtain the dispersion relation as a function of $\eta-\Delta$ only, i.e., the  nonlinear interaction strength ($\eta$) is such a linear term as photon-atoms detuning factor ($\Delta$). Fig. 2(a) shows such a shifting of dependencies.

By fixing a wave number $k=0.1$, in Fig. 2(b), the dependences of these two frequency branches are modified strongly with respect to the change of interaction strength $\eta$.
As the interaction energy among the qubits changes from a repulsive to an attractive one, both frequency $\omega_+$ and $\omega_-$  decrease as the value of the qubit interaction strength increases.
But when the above linear condition of $\eta-\Delta$ is not valid, these modified frequency curves for  two branches demonstrate a totally  different tendency to the frequency detuning, $\Delta$.
For a large enough negative (positive) value of the interaction strength, $\eta \ll -10$ $(\eta \gg 10) $, the upper (lower) branches for different detunings merge into a single curve.
That is, in this condition we have a frequency curve which is independent on the detunings.
Now a nonlinear interaction term $\eta$ is dominant.

An example of the  group velocity $v_{\pm}$  for two branches is shown in Fig. 2(c) as a function of qubit interaction strength $\eta$, for $k = 0.1$ under the condition (\ref{couplingcondition}) but with three different frequency detunings, i.e., $\Delta = -2$, $0$, and  $+2$, respectively.
It is noted that the group velocity in this work is normalized to $\alpha d$, instead of the speed of light.
Behaviors of the group velocity of light in the cavity array  under discussion can be easily understood by using a physical picture of polariton propagation -- see Sec.~IV.
Here we note, that  the difference between these group velocities in the two branches becomes larger while the absolute value of the qubit interaction strength $\eta$ increases.
A "slow" propagation of the wave-packets can be achieved just by changing the qubit interaction parameter $\eta$ for a negative (positive) value for the UB (LB) solution.

\section{Polariton soliton solutions}
Now, we do a multiple-scale expansion up to the third order by collecting terms with $\lambda ^{3} $, and reach a  complex GLE~\cite{Aranson2002}, i.e., 
\begin{equation}
\label{GLE}
i\frac{\partial \Psi }{\partial t} +D_{\pm } \frac{\partial ^{2} \Psi }{\partial X_{\pm }^{2} } +C_{\pm } \left|\Psi \right|^{2} \Psi =0,
\end{equation}
where variables $\Psi =\lambda E^{(1)} $, $X_{\pm } =\xi _{\pm } /\lambda $, and $t_{2} =\lambda ^{2} t$  have been introduced. The coefficients $D_{\pm}$ and $C_{\pm }$, appearing in the second and third terms of Eq.~(\ref{GLE}) respectively, have the forms,
\begin{eqnarray}
D_{\pm } &=&\frac{\alpha d^{2} \Omega _{\pm }^{3} +{\rm v}_{\pm }^{{\rm 2}} g^{2} }{\Omega _{\pm }^{} \left(\Omega _{\pm }^{2} +g^{2} \right)},\label{coeffD}\\
C_{\pm }&=&\frac{n_{\rm pol} g^{4} \left[\left(3\Omega _{\pm }^{} +\Omega _{\pm }^{*} \right) -4\eta \right]}{2\left|\Omega _{\pm }^{} \right|^{2} \left(\Omega _{\pm }^{2} +g^{2} \right)}.\label{coeffC}
\end{eqnarray}
Coefficients defined in Eqs. (\ref{coeffD},\ref{coeffC}) are complex and can be evaluated as $D_{\pm } =D_{\pm }^{(1)} +iD_{\pm }^{(2)} $, $C_{\pm } =C_{\pm }^{(1)} +iC_{\pm }^{(2)} $.
Real part $D_{\pm}^{(1)} $ of $D_{\pm } $-coefficient describes diffraction effects occurring with the wave-packet; while the imaginary part $D_{\pm }^{(2)}$ characterizes diffusion processes.
Parameter $C_{\pm }^{(1)}$ is responsible for a Kerr-like nonlinearity that occurs due to polariton-polariton scattering. At the same time an imaginary part, that is $C_{\pm }^{(2)}$, is relevant to nonlinear absorption effects.

By taking into account Eq.~(\ref{Eq:DispRel}), it is helpful to introduce convenient UB ($m_{+}$) and LB ($m_{-}$) polariton masses  
\begin{eqnarray}
\label{masses}
m_{\pm } =\hbar \left[\left. \partial _{k}^{2} \omega _{\pm } \right|_{k=0} \right]^{-1} =\frac{m_{ph} \left(\Omega _{\pm }^{2} +g^{2} \right)}{\Omega _{\pm }^{2} } ,
\end{eqnarray}
where $m_{ph} ={\hbar \mathord{\left/ {\vphantom {\hbar  2\alpha d^{2} }} \right. \kern-\nulldelimiterspace} 2\alpha d^{2}}$ is an effective photon mass in the cavity~\cite{Barinov}.
 Neglecting the kinetic energy of polaritons and supposing that the solutions are taken at the bottom of dispersion curve,  one can approximate  
\begin{equation}
D_{\pm } \simeq \frac{\hbar }{2m_{\pm } }, \qquad C_{\pm } \simeq \pm \frac{2 n_{\rm pol} gY_{\mp }^{3} }{Y_{\pm } }, 
\label{BottomCoeff}
\end{equation}
where we have introduced Hopfield coefficients such as ${Y_{\pm }^{} =\frac{1}{\sqrt{2} } \left[1\pm \frac{\delta }{\sqrt{\delta ^{2} +4g^{2} } } \right]^{1/2}}$ and for the sake of simplicity we  suppose that $g>0$. The latter parameters determine the contributions from photons and qubit excitations  to polaritons.
In particular, we can express field variable $\Psi$ that is a \textit{flight qubit} through \textit{polariton  qubit} states as following
\begin{eqnarray}
\label{PsiPsi}
\Psi =Y_{+} \Xi _{UB} -Y_{-} \Xi _{LB},
\end{eqnarray}
where  $\Xi _{UB} $ and $\Xi _{LB} $ are new variables  characterizing UB and LB polaritons respectively. They are defined in a convenient way by using Bogoliubov transformation~\cite{Sedov}. 

In particular, for a positive and large frequency detuning $\delta $, that is $\left|\delta \right| \gg g$ and $\delta >0$, we have $Y_{+} \approx 1$ and $Y_{-} \approx \frac{g}{\left|\delta \right|}$, respectively, which corresponds to \textit{photon-like UB  polaritons} ($\Xi_{UB} \simeq \Psi $) with mass  $m_{+} \approx m_{ph} $ and group velocity $v_{+} =\hbar k/m_{+}$, see Eqs. (\ref{masses}-\ref{PsiPsi}) and Fig.~2(c).
Remarkably, at the same time the group velocity of LB polaritons with mass $m_{-} \approx m_{ph} \frac{\delta ^{2} }{g^{2} }$ approaches $v_{-} =\hbar k/m_{-} \ll c$, where $c$ is speed of light in vacuum.
Now, we have a "slow'' (matter-like) soliton  formation in the cavity array.
For $\left|\delta \right| \gg g$ but  $\delta <0$, a physical picture becomes opposite and a propagating optical pulse   relevant to \textit{LB  polaritons} ($\Xi_{LB} \simeq -\Psi $) with mass $m_{-} \approx m_{ph} $ and group velocity $v_{-} =\hbar k/m_{-} \gg v_{+} $.
In the presence of  qubit-light resonance condition, $\delta =0$, UB and LB polaritons  equally contribute to the qubit  state, that is $Y_{\pm }^{} =1/\sqrt{2} $.  The masses of  polaritons are equal to each other, i.e.,  $m_{\rm pol} \equiv m_{\pm } =2m_{ph} $.  Flight qubit propagates in the medium with the velocity  $ v_{\pm } =\hbar k/m_{\rm pol} $.

At thermal equilibrium (or quasi-equilibrium), the lower polariton branch is macroscopically occupied. In this case we are interested in the properties of LB polaritons only. 
Taking into account the dependence of  $\Psi $ on  LB polariton variable $\Xi _{LB} $ in Eq. (\ref{PsiPsi}), it is easy to  rewrite the GLE in Eq.  (\ref{GLE}) for LB polariton variable as
\begin{equation}
\label{NLSE}
i\frac{\partial \Xi _{LB} }{\partial t} +\frac{\hbar }{2m_{-} } \frac{\partial ^{2} \Xi _{LB} }{\partial X_{-}^{2} } -C_{p} \left|\Xi _{LB} \right|^{2} \Xi _{LB} =0,
\end{equation} 
where parameter $C_{p} \simeq 2n_{\rm pol} gY_{+}^{3}/Y_{-}$ establishes two-body polariton-polariton interaction strength.
Thus, Eq.~({\ref{NLSE}) derived  above by MSEF method remains in complete agreement with the  results obtained for LB polaritons by using  Hamiltonian diagonalization approach in~\cite{Sedov}.

\begin{figure}
\includegraphics[width=8cm]{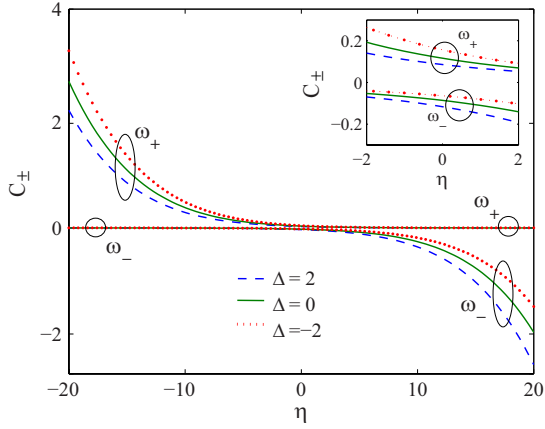}
\caption{(Color online) Nonlinear coefficient $C_{\pm}$ of the reduced NLSE in  Eq.~(\ref{GLE}) is shown for two frequency branches $\omega_\pm$, as a function of  the qubit interaction strength $\eta$ for different frequency detunings $\Delta$.
Inset shows a blowup close to $\eta = 0$. 
The parameters used are $k=0.1$, $\alpha=d=1$, $n_{\rm pol}=0.01$ and $g=10$, respectively.}
\end{figure}

It is useful to examine Eqs.~(\ref{GLE}-\ref{coeffC}) in the limiting (dissipationless) case for the system by  setting $\gamma _{c} = \Gamma _{d} =0$. In this limit, both coefficients $D_{\pm}$ and $C_{\pm}$ become real and GLE in Eq. (\ref{GLE}) is reduced to a standard nolinear Schr\"odinger equation (NLSE), that possesses exact \textit{bright} and \textit{dark} soliton solutions depending on the sign of the coefficients $C_{\pm} =C_{\pm}^{(1)}$ and $D_{\pm} =D_{\pm}^{(1)}$.
Here, we suppose that coefficients $D_{\pm } >0$. 
Then,  for the case of $C_{\pm } >0$, we have a \textit{bright }soliton solution; while for the case of $C_{\pm } <0$, \textit{dark} solitons can be found ~\cite{Agrawal1989}.

\begin{figure}
\centering
\includegraphics[width=8cm]{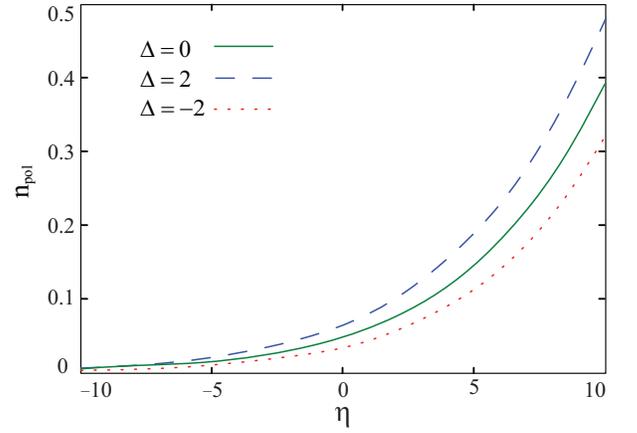}
\caption{(Color online) \label{polnumber} Polariton number density  $n_{\rm pol} $ versus  qubit interaction strength $\eta$  required for soliton formation taken for $\gamma = 0.1$, $\eta = 1$.  Other parameters are the same as in Fig.~3.}
\end{figure}

In Fig. 3, we show the dependences of this  nonlinear coefficient, $C_{\pm}^{(1)}$, as a function of the  qubit interaction strength $\eta$.
In Fig. 3, we see that the upper (optical) branch gives only a non-negative value of $C_{\pm}^{(1)}$; while the lower branch that characterizes matter excitations gives only a non-positive value.
In other words, we can only have a bright soliton solution supported in UB; while a dark one in LB. 
But due to the requirement to conserve the polartion number only bright soliton solutions can meet this condition for existing~\cite{YY-SIT}.
For a moderate polariton density number of total excitations, that is $n_{\rm pol} = 0.01$, as shown in the inset of Fig. 3, the value of $C_{\pm}^{(1)}$ tends to zero for the attractive qubit interaction, $\eta > 0$.
But to one's surprise, as compared to the positive one, a negative qubit interaction $\eta < 0$ can induce a larger nonlinear coefficient, $C_{\pm}^{(1)}$, which is enhanced by two orders of the magnitude.
At the same time, as shown in Fig. 2(c),  this bright soliton also moves as a "slow light". 
This giant enhanced nonlinearity is inherent in matter-like polaritons and results from the repulsive interaction among the qubits in this case~\cite{SichKrizhanovskii2012}.

The difference for the nonlinear coefficient $C_{+}$ between positive and negative values of $\eta$ can be understood in the following way.
For a positive value of $\eta$, the nonlinear interaction among qubits is attractive. In this region,  the induced nonlinearity mediated through qubit interactions is almost zero.
Instead, for a negative value of $\eta$, the repulsive qubit interaction effectively  contributes to a great amount of  the induced nonlinearity of the photon-like UB polariton soliton. 
Moreover, the properties of nonlinear coefficient $C_{\pm}^{(1)}$ being a function of polariton number density $n_{\rm pol}$ are evident from Eq.~(\ref{coeffC}).
Effectively, if one increases polariton number density $n_{\rm pol}$ at each cavity, the effect of the ensemble interacting qubits results in the enhancement value of a nonlinear coefficient. Practically, this effect can be used for creating artificial cluster materials based on physically small but macroscopical volumes of a qubit-light interaction. Photonic crystal defect cavities seem to be an appropriate candidate for providing such interaction, cf.~\cite{Sedov}.

\section{\label{polaritonsolitonsunderperturbation}BRIGHT POLARITON SOLITONS UNDER PERTURBATIONS}

Now we are going to examine a coupled qubit-light system in respect of bright soliton formation; the system being non-equilibrium. In this case we are interested in soliton solutions of Eq.~(\ref{GLE}) for optical field amplitudes $\Psi $ in the presence of polariton formation. It is useful to bring  Eq.~(\ref{GLE}) to a dimensionless form, rewriting it as follows
\begin{eqnarray}
i \frac{\partial \Psi }{\partial \tau} +\frac{1}{2} \frac{\partial ^{2} \Psi }{\partial x^{2} } +|\Psi |^{2} \Psi =
-i\varepsilon_{1} |\Psi |^{2} \Psi +i\varepsilon_{2} \frac{\partial ^{2} \Psi }{\partial x^{2} }, \label{GLEDimensionless}
\end{eqnarray}
where $x =X_{+ } /d$ and $\tau=\frac{2 D_{+ }^{(1)} }{d^{2}} t$ are new dimensionless spatial and temporal coordinates respectively; ${\varepsilon _{1} =\frac{C_{+}^{(2)} }{\left|C_{+}^{(1)} \right|} }$,  ${\varepsilon _{2} =\frac{|D_{+}^{(2)} |}{2D_{+}^{(1)} } }$ are perturbation coefficients ($\varepsilon_{1,2}>0$). In Eq.~(\ref{GLEDimensionless}) we also use the fact that the characteristic time scale of a dispersion action and the influence of nonlinearity on wave packet spreading should be the same~\cite{Agrawal1989}. In particular, the condition
\begin{equation}
\label{dencity}
n_{\rm pol} =\frac{4D_{+ }^{(1)} \left|\Omega _{+ }^{} \right|^{2} }{d^{2} g^{4} } \left|{\rm Re}\left[\frac{3\Omega _{+ }^{} +\Omega _{+ }^{*} -4\eta }{\Omega _{+ }^{2} +g^{2} } \right]\right|^{-1}
\end{equation} 
should be met for polariton number density $n_{\rm pol}$ in this case.

Dependences of $n_{\rm pol}$ as a function of qubit interaction strength $\eta$ are presented in Fig.~4.
The limiting cases for polaritons are determined by the magnitude and sign of detuning $\delta $ that includes the dependence on $\eta$. For large positive values of $\eta $, low-excitation density limit $(n_{\rm pol} \ll 1)$ is  violated for the chosen qubit-light interaction parameters. In this case we should  take into account  next terms in expansion of operators $\hat{J}_{+,i} $ characterizing a two-level oscillator system, resulting in a quintic nonlinearity ~\cite{Sedov}.

\begin{figure}
\centering
\includegraphics[width=8.4cm]{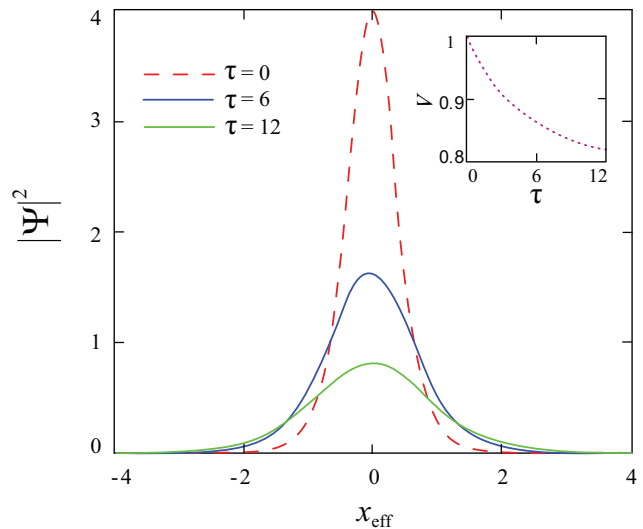}
\caption{(Color online) \label{bsolitonshape} Snapshots of bright soliton profile $\left|\Psi \right|^{2} $ as a function of effective coordinate $x_{\rm eff}=x-\zeta(\tau)$ at  different time of propagation  $\tau$. In the inset  temporal behavior of soliton group velocity $v$ is plotted. The parameters are (cf.~\cite{EgorovSkryabinLederer2010,KarrBaas2004}): $g=2 \pi \times 1.7 {\rm \, THz}$, $\Gamma=2 \pi \times 12.1 {\rm GHz}$, $\gamma_{c}=2 \pi \times 50 {\rm GHz}$, $\alpha=2 \pi \times 0.75 {\rm THz}$, $\eta n_{\rm pol}=2 \pi \times 24.3 {\rm GHz}$, $d= 400 {\rm nm}$, $k=10^{4} {\rm m^{-1}}$, and $\delta = 0$, respectively.}
\end{figure}

We examine the GLE in Eq.~(\ref{GLEDimensionless}) using perturbation theory for solitons by consulting the reference~\cite{KarpmanMaslov}.
In particular, we are looking for the solution of Eq.~(\ref{GLEDimensionless}) in the form
\begin{equation}
\label{BrightS}
\Psi (\tau, x)=2\nu {\rm sech}\left[{\rm 2}\nu \left( x-\zeta (\tau )\right)\right]e^{i\varphi (\tau, x)},
\end{equation} 
where $\nu $, $\zeta (\tau )$, and $\varphi (\tau, x)$ are related soliton amplitude, position, and phase, respectively. In the absence of perturbation (${\varepsilon _{1,2} =0}$),  the ansatz used in Eq.~(\ref{BrightS}) represents an exact solution for Eq. ~(\ref{GLEDimensionless}) with the parameters evolving in time as  
\begin{eqnarray}
\label{Phase}
\zeta (\tau) &=&\frac{v \tau }{2} +\zeta _{0},\\
\varphi (\tau, x) &=& \frac{v }{ 2} \left(x-\zeta (\tau )\right)+\delta (t),\\
\delta (\tau ) &=& \frac{v^{2} \tau }{2} +2\nu ^{2} \tau +\delta _{0}, 
\end{eqnarray}
where  parameters $\zeta _{0} $ and $\delta _{0} $ represent the initial soliton position and phase respectively. 
Here,  $v \equiv v_{+} $ is soliton velocity that we can associate with the velocity of UB polaritonic wave-packet by setting $\gamma =\Gamma _{d} =0$.
In the presence of small perturbation ($\varepsilon _{1,2} \ne 0$), we consider the soliton amplitude $\nu $ and velocity $v$ to be time dependent.
Applying  soliton perturbation theory, we arrive at the set of equations for solitons parameters under the adiabatic approximation,
\begin{eqnarray}
\label{munuzetadeltadot}
\dot{\nu } &=&-\frac{8}{3} \left(2\varepsilon _{1} + \varepsilon _{2} \right)\nu ^{3} -\frac{1}{2} \varepsilon _{2}  v^{2} \nu,\\
\dot{v}_{+} &=& -\frac{16}{3} \varepsilon _{2} v \nu ^{2},\\
\dot{\zeta } &=& \frac{v }{ 2}, \\
\dot{\delta } &=& \frac{v^{2} }{2} +2\nu ^{2},
\end{eqnarray}
where dot denotes derivative in respect of normalized time, i.e., $d/d\tau $.

In Fig.~5,  we represent a typical UB polariton soliton profile taken at different time for qubit-light interaction in a semiconductor GaAs-based  quantum well (QW) structure, as an example.
Based on real parameters of GaAs-based quantum wells, the initial shape of this soliton is characterized by the dashed (red) curve.

From Fig.~5, it is evident that a soliton amplitude vanishes due to effective nonlinear absorption. At the same time the group velocity of soliton is slightly  modified due to diffusion processes characterized by parameter $\varepsilon _{2} $ shown in  Eq.~(\ref{munuzetadeltadot}).
The characteristic time of disturbing soliton group velocity is tens of picoseconds which seems to be reasonable taking into account polariton lifetime in semiconductor structures (the maximal magnitude of dimensionless time  $\tau=12$  in the inset of Fig.~5 corresponds to the value of 25 picoseconds in real time scale),~cf.~\cite{SichKrizhanovskii2012,BarlandTredicce2002,AmoPigeon2011}. 
Moreover, the temporal soliton displacement $\zeta (\tau )$ can be characterized by Eq.~(\ref{Phase}) with a good approximation.

\section{\label{concl} Conclusions}

In this work,  we consider the problem of  soliton formation in the  array of weakly coupled qubit  ensembles interacting with the optical field in a tunnel-coupled cavity array. The effects of cavity field dissipation and qubit dephasing have been taken into account. By expanding Lipkin-Meshkov-Glick model for the ensemble of interacting qubits with the cavity photon fields in the array, we analytically reveal the existence of mean-field soliton solutions. In particular, we use multiple-scale envelope function method to obtain an effective complex Ginzburg-Landau equation, containing nonlinear absorption and diffusion terms.
Mean-field soliton solutions for this coupled matter-field states are revealed  in the  representation of  UB and LB polaritons.
With a negative value of the qubit interaction, not only a "slow" wave-packet in the form of solitons but also an enhanced nonlinearity can be achieved in such a qubit-cavity-QED array configuration.
We have shown that in this case an optical wave-packet can be recognized as a macroscopical polariton qubit state.
In particular, we have demonstrated that bright solitons are formed for UB polariton wave packets and characterized by slowly varying soliton amplitude, momentum, position, and phase. Notably, the group velocity of soliton decreases in time due to diffusion processes within a picoseconds domain for semiconductor structures.
Our analytical and numerical results presented here and connected with a cavity-QED array provide a platform for the studies of collective properties of light with interacting media.

\begin{acknowledgments}
This work was supported by RFBR Grants No. 10-02-13300, 11-02-97513, 12-02-90419, 12-02-97529 and by Russian Ministry of Education and Science under the contracts 14.740.11.0700. 
\end{acknowledgments}

\end{document}